\documentclass[a4paper,twoside]{article}
\newcommand{\affil}[1]{$^{\rm #1}$}
\usepackage[authoryear]{natbib}
\bibpunct{(}{)}{;}{a}{}{,}
\usepackage{graphicx}
\date{}

\title{\large\bf\flushleft Asymptotic Giant Branch models at very low metallicity}
\author{\parbox{\textwidth}{\flushleft
\vspace{-0.5cm} {\it S. Cristallo\affil{1,*}, L.
Piersanti\affil{1}, O. Straniero\affil{1}, R. Gallino\affil{2},
I. Dom\'inguez\affil{3} and F. K\"appeler\affil{4}}\\
\vspace{0.4cm}
{\small \affil{1}\,Osservatorio Astronomico
di Teramo (INAF), via Maggini 64100 Teramo, Italy}\\
{\small \affil{2}\,Dipartimento di Fisica Generale, Universit\'a
di Torino, via P. Giuria 1, 10125 Torino, Italy}\\
{\small \affil{3}\,Departamento de F\'isica Te\'orica y del Cosmos
, Universidad de Granada, 18071 Granada, Spain}\\
{\small \affil{4}\,Forschungszentrum Karlsruhe, Institut f\"ur Kernphysik Postfach 3460, D-76021 Karlsruhe, Germany}\\
{\small \affil{*}\,Email: cristallo@oa-teramo.inaf.it}}}
\begin{document}
\twocolumn[
\begin{changemargin}{.8cm}{.5cm}
\begin{minipage}{.9\textwidth}
\vspace{-1cm} \maketitle \small{\bf Abstract: In this paper we
present the evolution of a low mass model (initial mass
$M$=1.5 M$_\odot$) with a very low metal content ($Z=5\times
10^{-5}$, equivalent to [Fe/H]$=-2.44$). We find that, at the
beginning of the AGB phase, protons are ingested from the envelope
in the underlying convective shell generated by the first fully
developed thermal pulse. This peculiar phase is followed by a deep
third dredge up episode, which carries to the surface the freshly
synthesized $^{13}$C, $^{14}$N and $^{7}$Li. A standard TP-AGB
evolution, then, follows. During the proton ingestion
phase, a very high neutron density is attained and the s-process
is efficiently activated. We therefore adopt a nuclear network of
about 700 isotopes, linked by more than 1200 reactions, and we
couple it with the physical evolution of the model. We discuss in
detail the evolution of the surface chemical composition, starting
from the proton ingestion up to the end of the TP-AGB phase. }

\medskip{\bf Keywords:} stars: AGB and post-AGB --- physical data and processes:
nuclear reactions, nucleosynthesis, abundances

\medskip
\medskip
\end{minipage}
\end{changemargin}
]
\small

\section{Introduction}

The interpretation of the already available and still growing
amount of data concerning very metal-poor stars (see
\citet{bc05} for the classification of these objects) require
reliable stellar models. A significant fraction
($\sim$25\%) of stars with [Fe/H]$<$-2.5 show carbon-enriched
atmospheres (Carbon Enhanced Metal Poor stars - CEMP stars).
Moreover, many CEMP stars ($\sim$75\%) are characterized by
s-process enhanced patterns (CEMP-s): in this paper we will only
address to this sub-class of low metallicity stars. CEMP-s stars
belong to the Galactic Halo and, therefore, are old ($\sim 14$
Gyr) low mass ($M<0.9$ M$_\odot$) objects; notwithstanding, they
show the concomitant enhancements of C, N and s-process elements,
which are commonly ascribed to the complex interplay of
nucleosynthesis, mixing and mass loss taking place along the
Asymptotic Giant Branch (AGB). When these low mass stars reach the
AGB, their envelope is so small that third dredge up (TDU) may not
take place, so they are probably dwarfs or giants belonging to
binary systems. Therefore, the C-enhancement and the products of
the neutron capture nucleosynthesis could result from an ancient
accretion process (via stellar wind or, less likely, via
Roche lobe overflow) from the (now extinct) AGB companion
(see, e.g., \citealt{roe08,tho08}). Note that a recent
determination of radial velocities for a sample of CEMP-s stars
suggests that all of these objects are members of binary systems
\citep{luc}. Some of these stars present exotic chemical
distributions and, for
this reason, represent the current frontier of AGB modeling.\\
At very low metallicities, there is a minimum mass (which
increases with decreasing the metallicity) under which models
suffer proton ingestion from the envelope down to the underlying
convective He-intershell. This occurs during the first fully
developed thermal pulse (TP). We refer to this episode as the
Proton Ingestion Episode (PIE). Larger mass models follow a
standard TP-AGB evolution.  This feature has already been found by
many authors \citep{ho90,fu20,su04,iw04,stra04,cri07,cl08,lau09}: the
occurrence of PIE is therefore a robust prediction, in spite of
the different physics adopted in the various works. Moreover,
recent 3D hydrodynamical simulations \citep{wo08} confirm this
peculiarity, which characterizes very \begin{figure*}[t]
\begin{center}
\includegraphics[scale=0.55, angle=0]{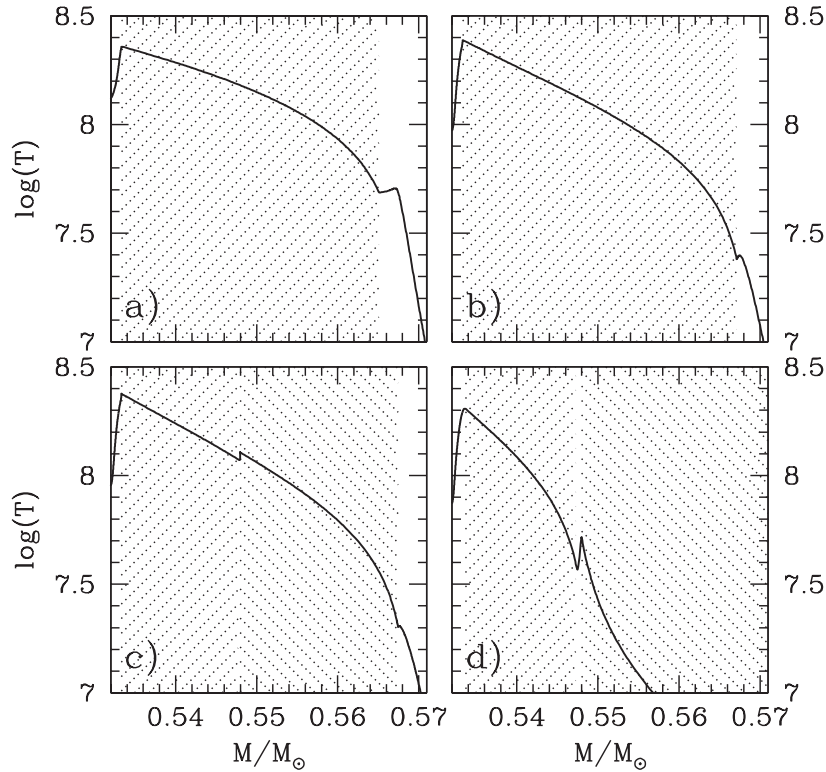}
\caption{Evolution of the temperature within the He-intershell
during the PIE. We plot the profiles at the beginning of proton
ingestion (panel \textbf{a}, $\Delta$t=0 yr), soon before the
convective shell splitting (panel \textbf{b}, $\Delta$t=1.6457
yr), soon after the convective shell splitting (panel \textbf{c},
$\Delta$t=1.6468 yr) and during the relaxing phase (panel
\textbf{d}, $\Delta$t=2.1843 yr). Note how the convective shell
(shaded region) in the top left panel is not yet fully
developed. Different shading directions (panel \textbf{c}
and \textbf{d}) indicate separate convective regions. See text
for details.}\label{temp}
\end{center}
\end{figure*}
metal-poor models. A full understanding of the involved physics is
therefore mandatory. \\In next Sections, we present and discuss
the evolution of an AGB model with initial mass $M$=1.5 M$_\odot$
and $Z=5\times 10^{-5}$, with emphasis on the effects on the
stellar structure caused by the PIE. The model has been computed
with the FRANEC stellar evolutionary code \citep{chi98};
details of the mixing scheme (see next Section) and
latest improvements to the code are described in \citet{cri09}. We
assume a solar-scaled initial distribution \citep{lodd}, although
at low metallicities a clear oxygen enhancement has been observed
by many
authors (see, e.g., \citealt{abia}).\\

\section{The Proton Ingestion Episode}\label{pie}

As outlined in the previous Section, at the beginning of the
thermally pulsing phase of low mass metal-poor models, the
convective shell powered by 3$\alpha$ reactions may extend up to
the base of the H-rich envelope. This only occurs at low
metallicities and is due to the flattening of the entropy peak
characterizing the H-burning shell. At low Z, in order to
compensate the reduction of CNO catalysts, the temperature of the
H-shell increases and, consequently, the entropy within the
H-shell decreases (entropy variation is defined as $dS=dQ/T$,
where dQ is the exchanged heat and T is the temperature).
Therefore, while at larger metallicities mixing is prevented by
the entropy peak of the H-shell \citep{iben}, at low metallicities
protons can be mixed within the convective shell and captured by
the abundant $^{12}$C. The temperature profile within the
convective shell runs from about $T= 5.0\times10^7$~K at the outer
border to $T= 2.3\times10^8$~K at the bottom (see top left panel
of Fig.~\ref{temp}). At such high
\begin{figure*}[t]
\begin{center}
\includegraphics[scale=0.5, angle=0]{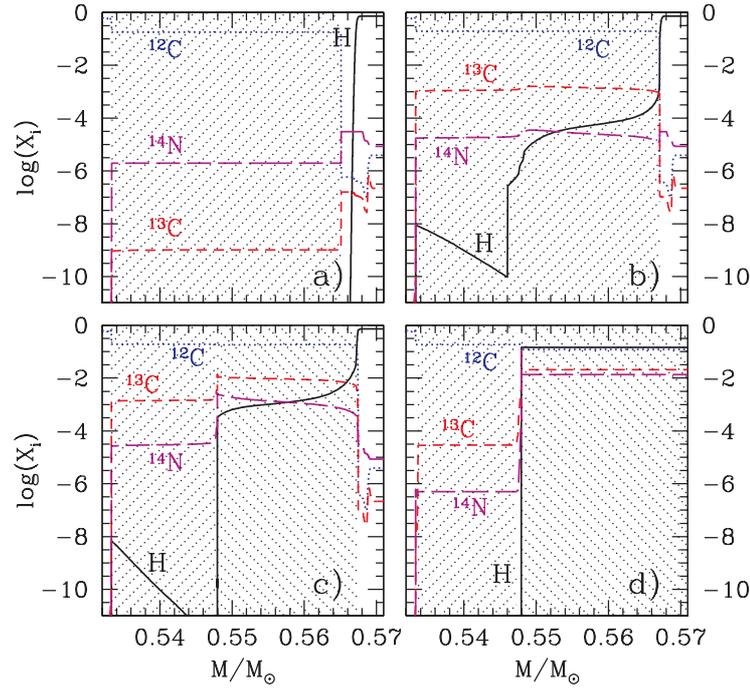}
\caption{Evolution of selected key isotopes within the He-intershell during the PIE.}\label{chimica}
\end{center}
\end{figure*}
temperatures H-burning is very efficient and, therefore, protons
are captured on-flight by CNO isotopes before reaching the bottom
of the convective shell. In these conditions, we can
distinguish three different regions within the flash-driven
convective shell:
\begin{itemize}
{\item the external zone, where the CNO-burning time scale
($\tau_{\rm CNO}$) is largely greater than the mixing time scale;}
{\item the inner zone, where temperature is so high that the
mixing efficiency is largely smaller than the burning one;} {\item
an intermediate zone, where the nuclear and mixing time scales are
comparable.}
\end{itemize}
The best approach to this situation is the simultaneous
solution of the differential equations describing the physical
structure of a star, the nuclear burning and the mixing. However,
it has to be noticed that this computational scheme can be used
only with small nuclear network, while it can not be applied when
using very large network, including several hundreds of isotopes
and nuclear processes. For this reason, in our simulations we
adopt a different approach. We fully couple the structure
equations with those describing the nuclear burning, so that we
directly take into account the energetic feedback of nuclear
reactions on the physical properties of the model (mainly the
temperature profile). Moreover, this allows us to determine
accurately the temperature gradient and, thus, the location and
the mass extension of the zones unstable for convection. Hence,
convection-induced mixing is computed by adopting the time
dependent formalism described in \citet{chi2001}, where the mixing
efficiency depends linearly on the temporal time step $\Delta t$
of the model. In our simulation, during the PIE, we limit $\Delta
t$ to 50\% of the mixing turnover time scale for the whole
convective shell ($\tau_{mix}$), in order to avoid the unrealistic
fully homogenization of that region. All the chemical species are
mixed within the whole unstable zone, with the exception of
protons, which are mixed from the outer border of the convective
shell down to the mass coordinate $M_P$, where $\tau_{CNO}=f
\Delta t$, $f < 1$ being a free parameter. According to our choice
on $\Delta t$, at $M_P$ $\tau_{CNO}$ results definitively smaller
than $\tau_{mix}$ (at least by a factor of $2/f$): this implies
that protons consumption via nuclear burning is much more rapid
than the supply via the convection-driven mixing. As a
consequence, in nature below this point it is hardly unlikely for
protons to survive. In the computation of PIE, we fix $f=1/3$
(this implies that at $M_P$ the temperature roughly corresponds to
$T\sim 1.3\times 10^8$). Such a choice is based on a a series of
evolutionary sequences computed with different values of $f$. Our
results show that for $f>1/3$ protons are accumulated in cool
zones where H-burning is inefficient, while for $f < 1/3$ protons
are mixed down to zones inconsistent with the on-flight burning.\\
In panel \textbf{a} of Fig.~\ref{chimica} we report the abundances
of H, $^{12}$C, $^{13}$C and $^{14}$N at the beginning of proton
ingestion. Note the large $^{12}$C abundance, due to the partial
He-burning inside the convective shell via
\begin{figure*}[t]
\begin{center}
\includegraphics[scale=0.5, angle=0]{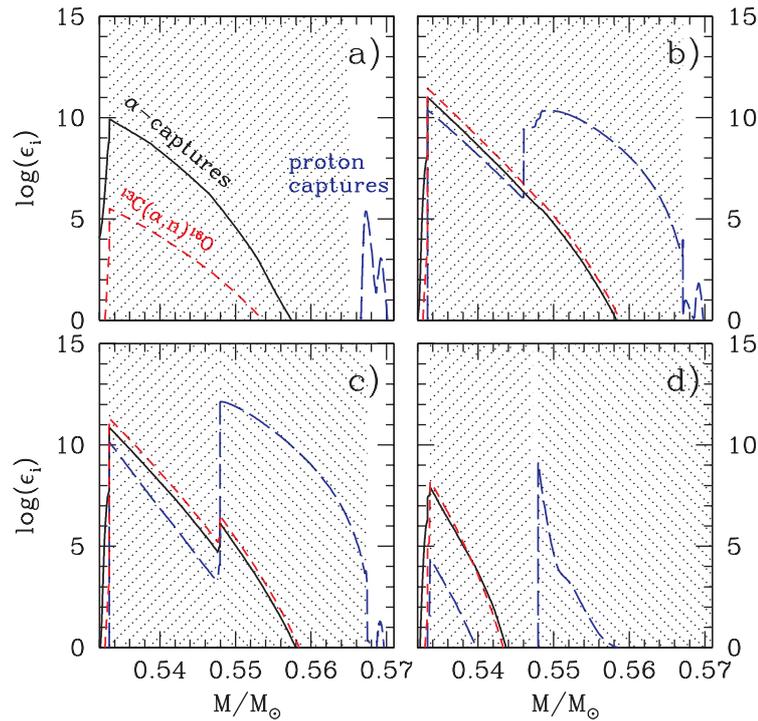}
\caption{Evolution of different contributions to energetics within
the He-intershell during the PIE.}\label{energetica}
\end{center}
\end{figure*}
3$\alpha$ reactions, and the relatively low abundances of $^{13}$C
and $^{14}$N, which result from the dilution, within the
convective shell, of the ashes left by the H-burning shell during
the previous interpulse phase. Similarly, in panel \textbf{a} of
Fig.~\ref{energetica} we plot the contributions to energetics
generated by proton captures (blue long-dashed curve) and by
$\alpha$ captures (dark solid curve). Two peaks clearly emerge
from the figure: the most internal one is due to 3$\alpha$
reactions (which power convection within the He-intershell) while
the second peak (more external and shallow) comes from the CNO
cycle in the H-burning shell. We plot apart the contribution of
the $^{13}$C($\alpha$,n)$^{16}$O reactions, in which we also
consider the energy release by the related neutron captures
(short-dashed red curve).
\\
Upper right panels (panels \textbf{b}) of Figures~\ref{temp},
\ref{chimica} and \ref{energetica} depict the situation 1.6457 yr
after the beginning of protons ingestion. Protons are captured by
the abundant $^{12}$C, leading to the nucleosynthesis of $^{13}$C
and $^{14}$N. Due to the concomitant actions of mixing and
burning, the resulting abundances strongly differ with respect to
the ones of the radiative H-burning shell, in which a full CNO
equilibrium is attained and the most abundant synthesized isotope
is $^{14}$N. During the PIE, the H-burning is instead always out
of equilibrium and this translates in a large production of
$^{13}$C and, at a lower level, of $^{14}$N, while $^{16}$O
remains practically untouched. Technically, this is due to the
fact that isotopes cannot locally pile up (and then reach the
equilibrium value) due to the presence of convective motions. Note
that $^{13}$C and $^{14}$N are correctly mixed down to the base of
the convective shell, because their burning timescales are always
lower than the model temporal step. The hydrogen profile extends
down to $M\sim 0.548$~M$_\odot$. The very small H abundance under
this limit basically derives from $^{14}$N(n,p)$^{14}$C reactions:
neutrons mainly come from the $^{13}$C($\alpha$,n)$^{16}$O
reaction, which is very efficiently activated at the bottom of the
convective shell. This is clearly shown in panel \textbf{b} of
Fig.~\ref{energetica}, in which emerges that
$^{13}$C($\alpha$,n)$^{16}$O reactions (and related neutron
captures) represent the major energy source at the bottom of the
convective shell. At the same time, the contribution of CNO
burning boosts a second energy peak, whose mass coordinate is
determined by the maximum penetration of protons within the
convective shell. When the maximum CNO energy peak slightly
exceeds the value of the He-burning peak, an inversion in the
temperature profile occurs (see panel \textbf{c} of
Fig.~\ref{temp}) and the convective shell splits into two
sub-shells: the lower one boosted by the
$^{13}$C($\alpha$,n)$^{16}$O reactions and the upper one by the
CNO cycle. After the splitting event, the two convective shells
follow separate evolutions.
\\
In the lower shell, the $^{13}$C($\alpha$,n)$^{16}$O reaction
consumes the available $^{13}$C and produces a very large neutron
flux.
\begin{figure*}[t]
\begin{center}
\includegraphics[width=9.5cm]{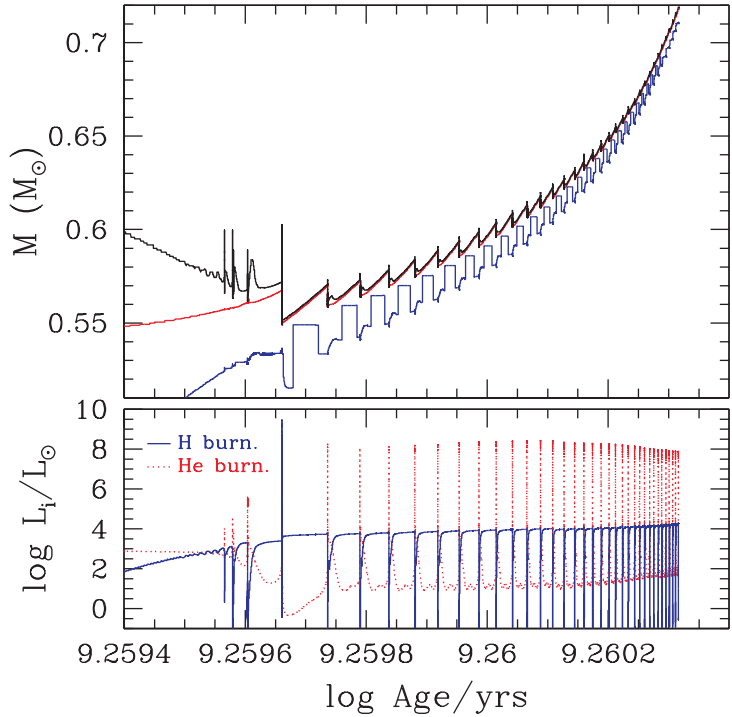}
\caption{Upper panel: temporal evolution of the mass coordinates
of (top to bottom) the inner border of the convective envelope,
the maximum energy production of the H-burning shell and the
maximum energy production within the H-depleted core. Note the
occurrence of a deep TDU (hereafter d-TDU) at the
beginning of the TP-AGB phase, as a consequence of proton
ingestion. Lower panel: temporal evolution of the
H-burning and He-burning contributions to
luminosity.}\label{mcore}
\end{center}
\end{figure*}
At the bottom of the shell, a neutron density as large as $n_n
\sim 10^{15}$ cm$^{-3}$ is attained, allowing the synthesis of
isotopes by-passed by a standard s-process. Since so high a
neutron density moves the nucleosynthetic path away from the
$\beta$ stability valley, we extended the nuclear network with
respect to that one used in previous works: the number of isotopes
is increased from 500 to about 700 and the number of nuclear
reactions grows from 700 to about 1200.
\\
In the upper shell, the $^{13}$C and the $^{14}$N continue to
increase thanks to the burning of freshly ingested
protons from the external layers. In this phase, the CNO cycle is
the major energy source in the star (see panel \textbf{c} of
Fig.~\ref{energetica}). Then, the structure starts expanding,
causing the temperature of the whole region to rapidly decrease
(see panel \textbf{d} of Fig.~\ref{temp}). During this phase, both
the energy sources boosting the two shells quench off (see panel
\textbf{d} of Fig.~\ref{energetica}); in the meanwhile, the upper
shell extends in mass.
\\
Later on, the envelope completely engulfs the upper shell,
dredging up to the surface a large amount of $^{13}$C and
$^{14}$N. The lower shell, thanks to the steep pressure and
density gradients formed in the splitting region, remains part of
the He-intershell region and the isotopes synthesized in this
region are later diluted in the following TP. Owing to the
increase of the CN abundance in the envelope, the further
evolution of this models resembles that of a standard AGB,
preventing the occurrence of further PIEs. In the upper
panel of Fig.~\ref{mcore} we report the temporal evolution of the
mass coordinates of (top to bottom): the inner border of the
convective envelope, the maximum energy production of the
H-burning shell and the maximum energy production within the
H-depleted core. In the lower panel we plot the H-burning and
He-burning contributions to luminosity. Note the H-burning burst
(L$_{\rm H}
>10^9 $L$_\odot$) caused by the PIE. This model undergoes 25
additional TDU episodes, each one followed by the formation of a $^{13}$C
pocket \citep{stra06,cri09}, which radiatively burns during the
interpulse period. The final surface composition is therefore a
combination of two different nucleosynthetic channels: the first
being a consequence of the PIE and the second coming from the
s-process in the radiative $^{13}$C pockets.
\\

\section{Nucleosynthesis Results}\label{resu}

\begin{figure*}[t]
\begin{center}
\includegraphics[scale=0.5, angle=0]{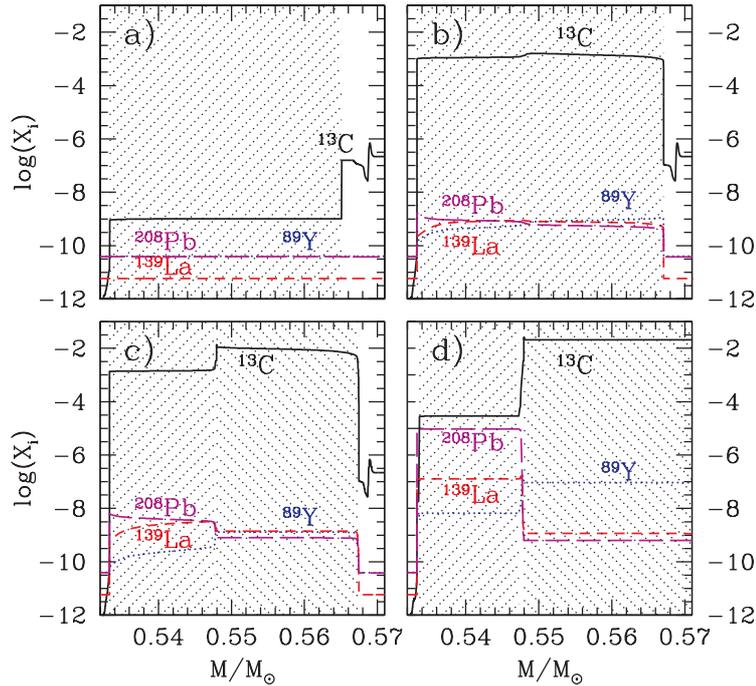}
\caption{Temporal evolution of selected heavy isotopes during the PIE.}\label{sproc}
\end{center}
\end{figure*}

\begin{figure*}[t]
\begin{center}
\includegraphics[scale=0.5, angle=0]{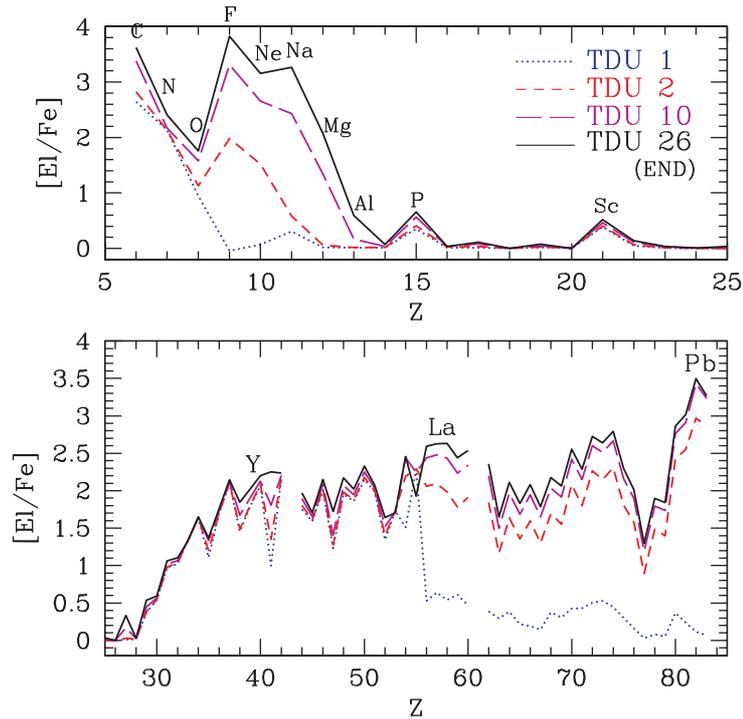}
\caption{Surface abundances (in spectroscopic notation) of a model
with initial mass M=1.5 M$_\odot$ model and $Z=5\times 10^{-5}$
after selected TDU episodes. Gaps at $Z=43$ and $Z=61$
correspond to unstable elements, Tc and Pm,
respectively.}\label{m1p5}
\end{center}
\end{figure*}

In Fig.~\ref{sproc} we report the evolution of selected heavy
isotopes during the PIE: $^{89}$Y (representative of light, ls,
s-process elements), $^{139}$La (representative of heavy, hs,
s-process elements) and $^{208}$Pb (which is strongly produced by
the s-process in metal-poor stars). Tracing s-process
elements have been selected basing on \citet{bu01}.
During the first phases of proton ingestion (see panel \textbf{a}
of Fig.~\ref{sproc}), the $^{13}$C abundance (about $10^{-9}$ in
mass fraction) does not allow an appreciable s-process production:
heavy elements abundances are therefore the initial ones (in the
previous phases of the stellar evolution no neutron sources have
been in fact efficiently activated). During the following phase,
up to the splitting event, we obtain a non negligible production
of heavy elements, with particular efficiency in the production of
hs elements (see panel \textbf{b} of Fig.~\ref{sproc}). Once the
splitting has occurred, the nucleosynthesis in the two shells
follow completely different behaviors. In the lower shell, where
$^{13}$C is destroyed at high temperatures ($T_{max}=2.5\times
10^8$~K), lead and hs elements are produced. In the upper shell,
instead, $^{13}$C burns at lower temperatures ($T_{max}\le
10^8$~K, see panel\textbf{ c} and \textbf{d} in Fig.\ref{temp})
and it is continuously added by the active proton burning. The
final effect is a net production of ls elements only (see panel
\textbf{d }in Fig.~\ref{sproc}). When the envelope dredges up the
isotopes synthesized in the upper shell, the surface distribution
of the model is therefore enriched in ls elements, showing hs and
lead abundances close to the initial ones (blue dotted curve in
Fig.~\ref{m1p5}). The mass dredged up during the d-TDU is
$\Delta M_{\rm TDU}\sim 1.7\times 10^{-2}$~M$_\odot$; the
corresponding efficiency (defined by means of the $\lambda$
factor\footnote{$\lambda$ is defined as the ratio between the mass
of the H-depleted material that is dredged up and the mass of the
material that has been burned by the H-shell during the previous
interpulse period.}) is 2.7. After the d-TDU episode, the model
follows a standard evolution. The isotopes synthesized in the
lower shell are diluted in the convective shell generated by the
following TP and their signature appears at the surface after the
second TDU episode (red short-dashed curve in Fig.~\ref{m1p5}).
The surface s-process distribution after the 2$^{nd}$ TDU mimics
the expected distribution of models at low metallicity
\citep{bi08}, characterized by a growing sequence ls-hs-lead. Such
a trend is even more strong as the number of TPs grows, displaying
a final distribution characterized by [ls/Fe]=2.02, [hs/Fe]=2.55
and [Pb/Fe]=3.50 (dark solid curve in Fig.~\ref{m1p5}). In the
plotted final distribution we took into account the decays of
radioactive isotopes (as, for example, the $\beta$ decay of
long-lived $^{93}$Zr to $^{93}$Nb) and, therefore, this
distribution can be compared with observational data of extrinsic
AGBs.
\subsection{Light elements}

\begin{figure}[t]
\begin{center}
\includegraphics[width=6.5cm]{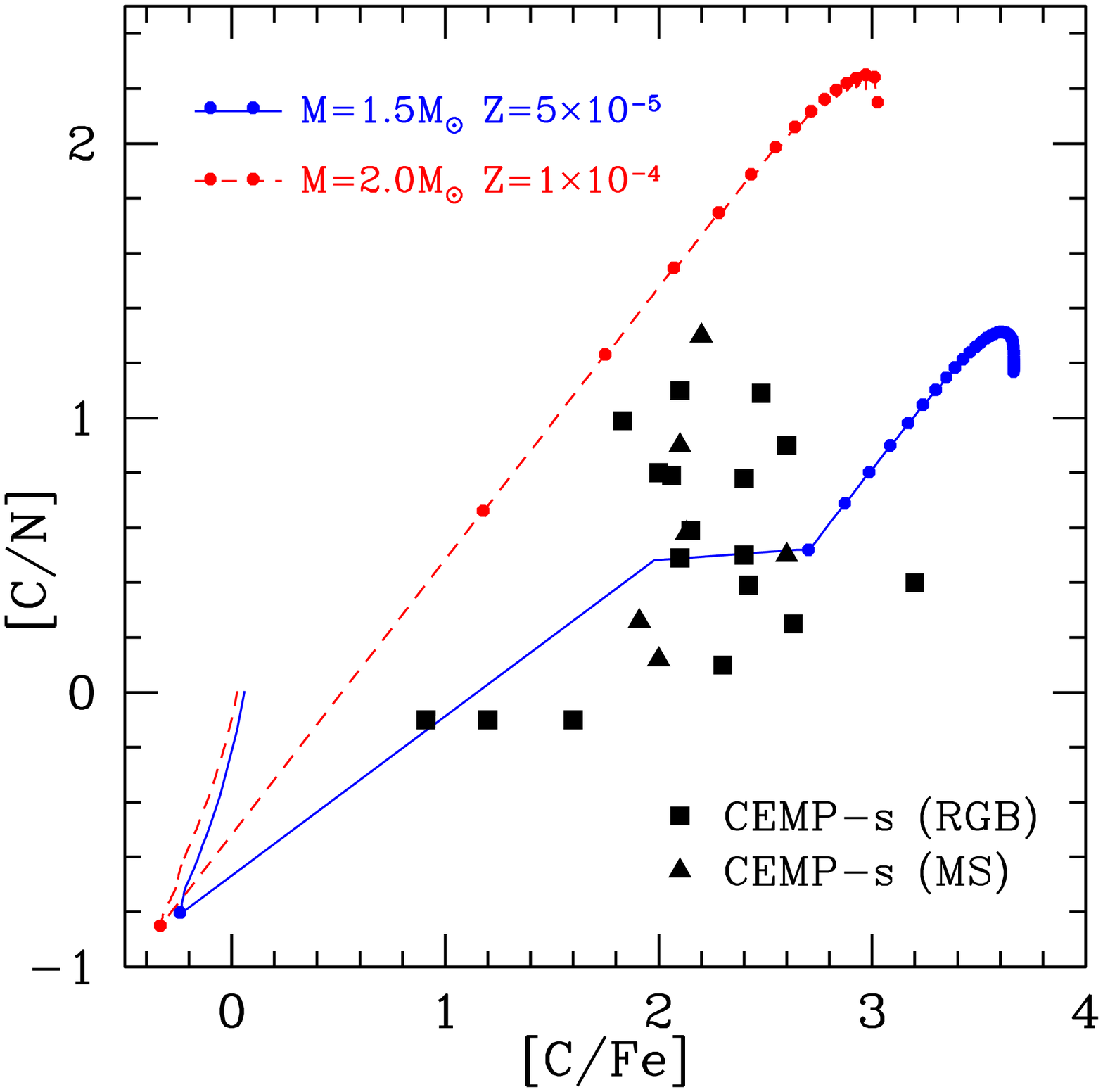}
\caption{[C/N] ratio as a function of [C/Fe] for the $M=1.5 $
M$_\odot$ model with $Z=5\times 10^{-5}$. We also report a model
with $M=2.0 $~M$_\odot$ and $Z=1\times 10^{-4}$. As a
comparison, we plot spectroscopic data of Giant and MS CEMP-s
stars. See text for details.}\label{light1}
\end{center}
\end{figure}

\begin{figure}[t]
\begin{center}
\includegraphics[width=6.5cm]{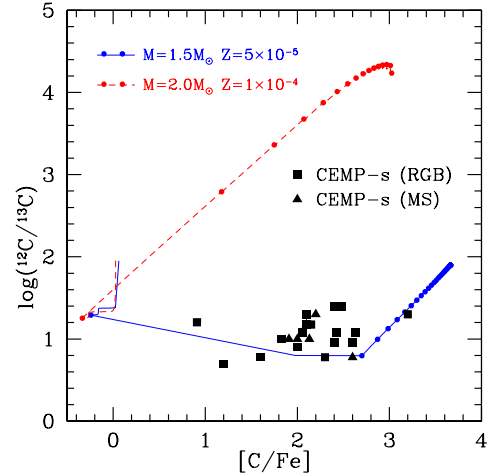}
\caption{$^{12}$C/$^{13}$C isotopic ratio as a function of [C/Fe]
for the $M=1.5 $ M$_\odot$ model with $Z=5\times 10^{-5}$. We also
report a model with $M=2.0 $~M$_\odot$ and $Z=1\times 10^{-4}$.
As a comparison, we plot spectroscopic data of Giant and
MS CEMP-s stars. See text for details.}\label{light2}
\end{center}
\end{figure}

As outlined in Section~\ref{pie}, a huge N enhancement results as
a consequence of proton ingestion from the top of the convective
shell, a possibility early recognized by \cite{ho90}. After the
d-TDU following the PIE, the surface is in fact enriched
not only in $^{12}$C (as in a standard TDU), but also in $^{13}$C
and $^{14}$N. In Fig.~\ref{light1} we plot the [C/N] ratio as a
function of [C/Fe], while in Fig.~\ref{light2} we report the
logarithm of the $^{12}$C/$^{13}$C isotopic ratio as a function of
[C/Fe]. In order to highlight the effects of the PIE, we
report the same ratios for a model with initial mass $M=2 $
M$_\odot$ and $Z=1\times 10^{-4}$ \citep{cri09}, which follows a
standard AGB evolution. Moreover, we also display
spectroscopic data of CEMP-s stars (both Giants and MS Stars)
extracted from the SAGA database \citep{saga}: our selecting
criteria have been metallicity (-2.9$<$[Fe/H]$<$-2.2), a
consistent s-elements enhancement ([Ba/Fe]$>$1.0) and data
completeness (values of C, N and $^{12}$C/$^{13}$C collected from
single surveys). The final [C/N] ratio attained by our 1.5
M$_\odot$ model better agrees with observational data of CEMP-s
stars, which show values ranging between 0 and 1.3. For the sake
of clarity, observational data show lower [C/Fe] ratios with
respect to our model, but since the majority of these
stars belong to binary systems \citep{luc}, a dilution factor has
to possibly be accounted for (depending on the masses and the
geometry of the system) \citep{ao08,tho08}. Moreover, other
physical mechanisms, such as gravitational settling, thermohaline
mixing and the occurrence of the First Dredge Up, eventually led
to a further dilution (\citet{stagle} and references therein).
Most of CEMP-s stars also show very low $^{12}$C/$^{13}$C
ratios (for some of them even close to the equilibrium value),
therefore proving that the material within their envelopes
experienced some H-burning. Our final value ($\sim 80$) is
strongly reduced with respect to models without the PIE (in which the
$^{12}$C/$^{13}$C ratio is larger than 10000) and, therefore,
at low metallicities PIE help explaining the low observed
carbon isotopic ratios. Observational data reported in
Fig.~\ref{light2} confirm this trend. Note that masses lower than
1.5 M$_\odot$ experience fewer TDU episodes and, therefore, lower
final $^{12}$C/$^{13}$C are expected with respect to the model
presented in this paper (Cristallo et al., in preparation).
Moreover, we remember that large enough masses (the exact value
depending on the metallicity) experience the Hot Bottom Burning
(HBB) \citep{sugi}, showing therefore $^{13}$C and $^{14}$N
enriched surfaces. However, the PIE can not occur at metallicities
larger than [Fe/H]$>-2$ \citep{cri07} and the mass experiencing
HBB grows when increasing the metallicity. Additional extra-mixing
mechanism, working during both the RGB and the AGB phase, are
therefore requested to reproduce the low $^{12}$C/$^{13}$C ratios
detected in stars of intermediate metallicity (see
\citealt{leb08}) as well in Galactic Open and Globular Clusters
(see, e.g., \citealt{gra}).
\\
Finally, let us discuss a further interesting feature, already
noted by \citet{iw04}, marking the PIE: a very large production of
$^{7}$Li. We obtain a final log$(\epsilon({\rm Li}))=3.74$, about
two order of magnitude higher than the lithium plateau observed at
low metallicity \citep{spsp}. The $^{7}$Li production mechanism is
very simple: during proton ingestion, $^{3}$He nuclei are ingested
within the convective shell (note that the envelope presents a
quite large $^{3}$He abundance, $X(^{3}{\rm He})\sim 3\times
10^{-4}$, due to the occurrence of the First Dredge Up). During
the PIE, $^{3}$He is rapidly captured by the abundant $^{4}$He,
leading to the synthesis of $^{7}$Be. This isotope then captures
an electron and synthesizes $^{7}$Li. At high temperatures,
lithium has a very large cross section against proton captures
and, if created in an hot environment, is rapidly destroyed.
However, given the rapidity of the stellar evolution after the
splitting event, the freshly synthesized beryllium has not enough
time to capture electrons in a hot environment and, therefore, is
rapidly diluted in the envelope by the d-TDU.
\section{Conclusions}
In this contribution we present the evolution of a low mass model
($M$=1.5 M$_\odot$) of very-low metallicity ($Z=5\times 10^{-5}$).
We discussed in details the effect that proton ingestion has on
the physical stellar structure and on the following
nucleosynthesis. As a by-product of the PIE, we found a large
production of $^{13}$C, $^{14}$N and $^{7}$Li. Moreover, we
discussed the surface s-process distributions after the deep TDU
following the PIE and after the standard TDU episodes. We are now
exploring the effects of the Proton Ingestion Episode on models
with different masses ($0.85\le M/M_\odot\le2.0$) and $Z=5\times
10^{-5}$ (Cristallo et al., in preparation). Proton
ingestion occurs in all these models. While in the 1.5 and 2.0
M$_\odot$ mass models the effects of proton ingestion on the final
surface compositions are integrated by the occurrence of standard
TDU episodes (as shown in this contribution), lower masses (0.85
and 1.0 M$_\odot$) show peculiar final surface compositions,
determined by the PIE only. Note that PIE remarkably lowers the
minimum mass for the occurrence of TDU.

\section{Acknowledgments}
S.C., O.S. and R.G. are supported by the Italian MIUR-PRIN 2006
Project ``Final Phases of Stellar Evolution, Nucleosynthesis in
Supernovae, AGB stars, Planetary Nebulae``. The authors
thank the anonymous referee, whose suggestions greatly improved
the quality of this paper.

\end{document}